\documentclass[
    final,
    5p,
    twocolumn,
    authoryear
]{elsarticle}

\date{July 19, 2023}

\journal{Automatica}

\usepackage[T1]{fontenc}
\usepackage[utf8]{inputenc}

\usepackage{booktabs}

\usepackage{amssymb,amsmath,mathtools}
\usepackage{units}
\usepackage{xspace}

\usepackage[nocompress]{cite}

\usepackage{csquotes}

\usepackage{tikz}
\usepackage{pgfplots}
\pgfplotsset{compat=1.9}
\usepgfplotslibrary{statistics}
\usetikzlibrary{arrows}
\usetikzlibrary{decorations.pathreplacing,decorations.markings}
\usetikzlibrary{calc}

\tikzset{
	on each segment/.style={
		decorate,
		decoration={
			show path construction,
			moveto code={},
			lineto code={
				\path [#1]
				(\tikzinputsegmentfirst) -- (\tikzinputsegmentlast);
			},
			curveto code={
				\path [#1] (\tikzinputsegmentfirst)
				.. controls
				(\tikzinputsegmentsupporta) and (\tikzinputsegmentsupportb)
				..
				(\tikzinputsegmentlast);
			},
			closepath code={
				\path [#1]
				(\tikzinputsegmentfirst) -- (\tikzinputsegmentlast);
			}
		}
	}
}

\tikzset{
	mid arrow/.style={
		postaction={
			decorate,decoration={
				markings,
				mark=at position .5 with {\arrow[#1]{stealth}}
			}
		}
	}
}

\usepackage{color}
\definecolor{plecs1}{rgb}{0,.7969,0}
\definecolor{plecs2}{rgb}{1,0,0}
\definecolor{plecs3}{rgb}{0,0,1}
\definecolor{plecs4}{rgb}{1,.7969,0}
\definecolor{plecs5}{rgb}{1,0,1}
\definecolor{plecs6}{rgb}{0,1,1}

\definecolor{vgRed}{RGB}{193, 48, 24}
\definecolor{vgOrange}{RGB}{243, 111, 19}
\definecolor{vgYellow}{RGB}{235, 203, 56}
\definecolor{vgGreen}{RGB}{162, 185, 105}
\definecolor{vgLightBlue}{RGB}{13, 149, 188}
\definecolor{vgDarkBlue}{RGB}{6, 56, 81}

\pgfplotsset{every axis/.append style={semithick,tick style={major tick
			length=4pt,semithick,black}}}

\pgfkeys{/pgfplots/x axis shift down/.style={
		x axis line style={yshift=-#1},
		xtick style={yshift=-#1},
		tick align=outside,
		xticklabel shift={#1}}}

\pgfkeys{/pgfplots/y axis shift left/.style={
		y axis line style={xshift=-#1},
		ytick style={xshift=-#1},
		yticklabel shift={#1},
		scaled y ticks = false,
    	tick align=outside,
		y tick label style={/pgf/number format/fixed,
		}
	}
}

\pgfplotsset{myPlot/.style={%
		width=8cm,
		height=3.5cm,
    	line width = 0.7pt,
		separate axis lines,
		axis x line*=bottom,
		x axis shift down = 3pt,
		enlarge x limits=false,
		axis y line*=left,
		y axis shift left = 6pt,
		enlarge y limits={abs=.25pt},
		enlarge x limits={abs=.25pt},
	}
}

\pgfplotsset{myBoxPlot/.style={%
    myPlot,
    clip=false,
    line width = 0.5pt,
    width= 7.8cm,
    height = 2.5cm,
    hide y axis,
    every boxplot/.style={mark=x, every mark/.append style={mark size=1.0pt}, draw=vgDarkBlue},
	}
}

\pgfplotsset{
      boxplot/draw/median/.code={%
        \draw[mark size=1.5pt,/pgfplots/boxplot/every median/.try]
          \pgfextra
          \pgftransformshift{
            \pgfplotsboxplotpointabbox
              {\pgfplotsboxplotvalue{median}}
              {0.5}
          }
          \pgfsetfillcolor{vgLightBlue}
          \pgfuseplotmark{*}
          \endpgfextra
        ;
      },
  }

\pgfplotsset{linestyle boxplot/.style={%
  boxplot = {%
    every box/.style={draw=none, fill=none},
    whisker extend=0,
  },
  mark=*,
  every mark/.append style={mark size=0.7pt, line width=0pt, opacity=0.6, fill=vgLightBlue}, draw=vgLightBlue,
  }
}

\pgfplotsset{resultsPlot/.style={%
		myPlot,
		width= 8cm,
		height = 4cm,
		xmin = 0,
		xmax = 336,
		xtick = {0, 48, 96, ..., 336},
		xticklabels = {0, 1, 2, ..., 7},
		clip=false,
		legend columns=4,  line width = .6pt,
		legend style={
			at={(1, 1)},
			yshift = .6,
			anchor=south east,
			draw=none,
			fill=none,
			legend cell align=left,
			/tikz/every even column/.append style={column sep=0.3cm}
		},
	}
}

\pgfplotsset{resultsPlot Power/.style={%
		resultsPlot,
		width = 8cm,
		height = 3cm,
		ymin = -2,
		ymax = 2,
		ylabel = {Power in \unit{pu}},
		font=\scriptsize,
	}
}

\pgfplotsset{resultsPlot powerFlow/.style={%
		resultsPlot,
		width = 3.5cm,
		height = 2.5cm,
		ymin = -1,
		ymax = 1,
		ytick = {-1, 0, 1},
		xlabel = {Time in \unit{d}},
		font=\scriptsize,
	}
}

\pgfplotsset{resultsPlot Energy/.style={%
		resultsPlot,
		width = 3.5cm,
		height = 2.5cm,
		ymin = -0,
		ymax = 6,
		ytick = {0, 3, 6},
		ylabel = {Energy},
		font=\scriptsize,
	}
}

\colorlet{colorPowerFlow}{cyan}
\colorlet{colorThermal}{plecs3}
\colorlet{colorGrid}{plecs2}
\colorlet{colorStorage}{orange}
\colorlet{colorRes}{plecs1}
\colorlet{colorDemand}{white!50!black}

\pgfplotsset{linestylePower/.style = {tension=0.01, thin,}}
\pgfplotsset{linestylePowerFlow/.style = {linestylePower, color=colorPowerFlow}}

\pgfplotsset{linestyleThermal/.style = {linestylePower, color=colorThermal}}
\pgfplotsset{linestyleGrid/.style = {linestylePower, color=colorGrid}}
\pgfplotsset{linestyleStorageEnergy/.style = {tension=0.01, thin, color=colorStorage, dash pattern = on 1pt off 1pt on 3pt off 1pt}}
\pgfplotsset{linestyleStoragePower/.style = {linestylePower, color= colorStorage}}
\pgfplotsset{linestyleRes/.style = {linestylePower, color= colorRes}}
\pgfplotsset{linestyleDemand/.style = {linestylePower, color= colorDemand}}
\pgfplotsset{linestyleMaximumRes/.style = {linestylePower, color=black!60!white}}
\pgfplotsset{linestyleLimit/.style = {const plot, color=black, dash pattern=on 1pt off 3pt on 3pt off 3pt, thin}}

\pgfplotsset{linestyleThermal stacked/.style = {linestylePower, color=colorThermal, fill=colorThermal!20!white}}
\pgfplotsset{linestyleGrid stacked/.style = {linestylePower, color=colorGrid, fill=colorGrid!20!white}}
\pgfplotsset{linestyleStorageEnergy stacked/.style = {color=colorStorage, dash pattern = on 1pt off 1pt on 3pt off 1pt}}
\pgfplotsset{linestyleStoragePower stacked/.style = {linestylePower, color=colorStorage, fill=colorStorage!20!white}}
\pgfplotsset{linestyleRes stacked/.style = {linestylePower, color=colorRes, fill=colorRes!20!white}}
\pgfplotsset{linestyleDemand stacked/.style = {linestylePower, color=colorDemand, fill=colorDemand!20!white}}

\newcommand{\plotLimit}[1]{%
	\addplot[linestyleLimit, forget plot] plot coordinates{
		(0,#1)
		(336,#1)
	};
}

\usepackage{todonotes}
\makeatletter
\renewcommand{\todo}[2][]{\tikzexternaldisable\@todo[#1]{#2}\tikzexternalenable}
\renewcommand{\missingfigure}[2][]{\tikzexternaldisable\@missingfigure[#1]{#2}\tikzexternalenable}
\newcommand{\atAj}[1]{\tikzexternaldisable\@todo[inline, color=blue!20]{@Ajay: #1}\tikzexternalenable}
\newcommand{\atCh}[1]{\tikzexternaldisable\@todo[inline, color=red!20]{@Chris: #1}\tikzexternalenable}
\newcommand{\fromAj}[1]{\tikzexternaldisable\@todo[inline, color=red!20]{Ajay says: #1}\tikzexternalenable}
\makeatother

\usepackage{acronym}
\acrodef{mg}[MG]{microgrid}
\acrodef{mpc}[MPC]{model predictive control}
\acrodef{res}[RES]{renewable energy sources}
\acrodef{al}[AL]{augmented Lagrangian}
\acrodef{dg}[DU]{distributed unit}
\acrodef{pcc}[PCC]{point of common coupling}
\acrodef{miqcp}[MIQCP]{mixed-integer quadratically-constrained program}
\acrodef{cc}[CC]{conditional cooperation}
\acrodef{fd}[FD]{feasible decomposition}
\acrodef{ac}[AC]{alternating current}\acused{ac}

\def\R{\mathbb{R}}

\def\N{\mathbb{N}}

\newcommand{\tran}{^{\mathstrut\scriptscriptstyle\top}}

\newcommand{\smallSum}{\textstyle\sum}
\newcommand{\diag}[1]{\ensuremath{\operatorname{diag}(#1)}}
\newcommand{\ok}{(k)}

\newcommand{\okhk}{(k + h | k)}

\newcommand{\boldSwitch}{\boldsymbol{\delta}}

\newcommand{\set}[1]{\{#1\}}

\def\vec{\operatorname{vec}}
\def\diag{\operatorname{diag}}

\def\E{\mathcal{E}}

\usepackage{algorithm}

\usepackage[breaklinks=true]{hyperref}
\usepackage{cleveref}

\crefname{figure}{Fig.}{Figs.}
\crefname{table}{Table}{Tables}
\crefname{section}{Section}{Sections}
\crefname{assumption}{Assumption}{Assumptions}
\crefname{algorithm}{Algorithm}{Algorithms}
\crefname{problem}{Problem}{Problems}
\crefname{proposition}{Proposition}{Propositions}
\crefname{remark}{Remark}{Remarks}
\Crefname{remark}{Remark}{Remarks}
\crefname{corollary}{Corollary}{Corollaries}

\newtheorem{remark}{Remark}
\newtheorem{corollary}{Corollary}

\newtheorem{assumption}{Assumption}
\newtheorem{proposition}{Proposition}
\newtheorem{problem}{Problem}
\newtheorem{proof}{Proof}
\renewcommand{\proof}[1][]{\ifthenelse{\equal{#1}{}}{\noindent {\itshape Proof:~}}{\noindent{\itshape Proof #1:}}\@\xspace}

\usepackage[inline, shortlabels]{enumitem}

\begin{document}

\begin{frontmatter}

\title{Distributed Conditional Cooperation Model Predictive Control~of~Interconnected~Microgrids\tnoteref{support}}
\tnotetext[support]{This work was partially supported by the German Federal Ministry for Economic Affairs and Energy (BMWi), Project No. 0325713A.}

\author[centrica]{Ajay Kumar Sampathirao}\ead{ajaykumarsampath@gmail.com}
\author[tub]{Steffen Hofmann}\ead{hofmann@control.tu-berlin.de}
\author[tub]{J\"{o}rg Raisch}\ead{raisch@control.tu-berlin.de}
\author[kassel]{Christian A. Hans}\ead{hans@uni-kassel.de}

\affiliation[centrica]{
    organization={Centrica Business Solutions},
    addressline={Roderveldlaan 2/b2},
    city={Antwerp},
    postcode={2600},
    country={Belgium}
    }
\affiliation[tub]{
    organization={Technische Universität Berlin},
    addressline={Einsteinufer 11},
    city={10587 Berlin},
    country={Germany}
    }
\affiliation[kassel]{
    organization={University of Kassel},
    addressline={Wilhelmshöher Allee 71--73},
    city={34121 Kassel},
    country={Germany}
    }

\begin{abstract}
    In this paper, we propose a model predictive control based operation strategy
    that allows for power exchange between interconnected microgrids.
    Particularly, the approach ensures that each microgrid benefits from power exchange with others.
    This is realised by including a condition which is based on the islanded operation cost.
    The overall model predictive control problem is posed as a mixed-integer quadratically-constrained program and solved using a distributed algorithm that iteratively updates continuous and integer variables.
    For this algorithm, termination, feasibility and computational properties are discussed.
    The performance and the computational benefits of the proposed strategy are highlighted in an illustrative case study.
\end{abstract}

\end{frontmatter}


\section{Introduction}
\label{sec:introduction}
Recent advances in renewable energies and growing concerns about environmental impacts of fossil-fuelled power plants led to a worldwide increase of \ac{res}.
\ac{res} are often small-scale units which are characterised by intermittent power output and geo\-graph\-i\-cal proximity to consumers.
At this juncture, the \acf{mg} concept is a promising approach to facilitate the integration of a large number of \ac{res}.
An \ac{mg} refers to a self-contained system with local demand, generators and storage units.
\acp{mg} can operate connected to or isolated from the power network, i.e., islanded.
For both modes, operation control serves to coordinate all units on a timescale of minutes.

Connecting \acp{mg} and facilitating power exchange between them can increase flexibility compared to islanded operation~\citep{ShaLiBah+17}.
The interconnection allows the \acp{mg} to benefit from different infeed pattern of geographically distributed and technologically diverse \ac{res}.
In this paper, we develop a strategy that manages power exchange while preserving the self-interest of each \ac{mg} in a network.
The derived conditional cooperation control scheme is composed of two parts:
\begin{enumerate*}[(i)]
	\item estimation of a condition for each \ac{mg} that reflects its self-interest; and\label{it:introduction:self-interest}
	\item determination of power exchange between the \acp{mg} in the network such that the condition from \ref{it:introduction:self-interest} is satisfied for each \ac{mg}.
\end{enumerate*}
The scheme is based on \acf{mpc} which is widely adopted for the operation of \acp{mg} as it allows to explicitly consider unit limits as well as economic objectives \citep{OuaDagDes+15}.

Coordinating multiple \acp{mg} in a network has been widely discussed in literature, e.g., by \citet{OuaDagDes+15, WanMaoNel15, SanSir17}.
They all highlight the benefits of coordinated power transfers in interconnected \acp{mg}.
However, they focus on minimising the combined operating cost which does not always secure the self-interest of each individual \ac{mg}.
This drawback has been dealt with by \citet{ParWieKyn+17,SCP2019}.
Unfortunately, their approaches require a central unit and are therefore not suitable for a fully distributed implementation.
Distributed \ac{mpc} that considers cooperation of individual agents while prioritising local goals has been studied, e.g., by \citet{VeRaWr05,VaEsBa+11}.
\citet{VaEsBa+11} formulated a feasible-cooperation \ac{mpc} as a cooperative game that allows each agent to decide whether it cooperates or not.
\citet{VeRaWr05} proposed a cooperation-based distributed \ac{mpc} approach, which iteratively exchanges information between the agents to find feasible \mbox{control actions}.

The main limitation of most approaches is their restriction to convex functions.
In the context of \ac{mg} operation control, this means that the on/off condition of conventional generators cannot be directly implemented.
This limitation was addressed by \citet{HanBraRai17}, where a hierarchical distributed \ac{mpc} that includes the switching of conventional units was proposed.
The binaries that represent their on/off condition were relaxed and power flow was decided using distributed optimization.
However, the approach did not preserve the \acp{mg}' self-interest.

We aim to fill this gap by introducing a novel \ac{mpc} approach.
Our main contributions are as follows.
\begin{enumerate*}[(i)]
	\item A \acl{cc} \ac{mpc} problem for the operation of interconnected \acp{mg} is posed as a \ac{miqcp}.
	It includes a condition that, contrary to \citet{HanBraRai17}, preserves each \ac{mg}'s self-interest, based on the cost of islanded operation.
  	\item We propose a feasible-decomposition-based algorithm to find a (not necessarily optimal) solution to aforementioned \ac{miqcp} which
  	comes with a significant reduction in computational complexity.
  	We study theoretical properties of the algorithm and provide a termination criterion.
	\item A fully distributed version of aforementioned algorithm is proposed.
	In contrast to \citet{HanBraRai17}, power flows are determined using a distributed \acl{al} method which does not require central coordinator.
	\item We compare our novel approaches with the ones proposed by \citet{HanBraRai17} in an extensive case study.
	In detail, we examine computational properties of the distributed approach and highlight its applicability.
\end{enumerate*}

The remainder of the paper is structured as follows.
In \cref{sec:interConnectedMg}, \ac{mg} and network model are introduced.
Then, in \cref{sec:optimalControlProblem} a \acl{cc} \ac{mpc} problem is derived.
In \cref{sec:outerApproximation}, an algorithm to solve this \ac{mpc} problem is deduced.
In \cref{sec:distributedOptimisation}, a distributed implementation of the algorithm is discussed.
Finally, in \cref{sec:caseStudy} a case study is presented.

\subsection{Notation and mathematical preliminaries}
\label{sec:preliminaries}
The set of real numbers is $\R$, the set of nonpositive real numbers $\R_{\leq 0}$, the set of nonnegative real numbers $\R_{\geq 0}$, the set of positive real numbers $\R_{>0}$ and
the set of negative real numbers $\R_{< 0}$.
The set of positive integers is $\mathbb{N}$,
the set of nonnegative integers $\mathbb{N}^0$ and
the set of the first $n$ positive integers $\N_{n} = \{1, 2, \ldots, n\}$.
Moreover, $\N_{[0, n]} = \{0\} \cup \N_{n}$.
The cardinality of a set $\mathbb{A}$ is $|\mathbb{A}|$.
The vector $x = [x_{v_1} ~ x_{v_2} ~ \cdots ~ x_{v_n}]$ with $v_i \in V = \{v_1, v_2, \cdots, v_n \} \subset \N$ and $v_i < v_j$ for $i < j$ is denoted by $\vec(x_{v_i})_{v_i \in V}$.
Moreover, $\vec(x_{v_iu_j})_{v_i \in V, u_j \in U}$ is shorthand for $\vec(\vec(x_{v_iu_i})_{u_i \in U})_{v_i \in V}$.
The diagonal matrix with entries $x_{v_1}, x_{v_2}, \cdots, x_{v_n}$ is denoted by $\diag(x)$.
Finally, $\| \cdot \|$ is the $L_{2}$ norm and $|c|$ the absolute value of $c\in\R$.

Consider an undirected connected graph $\mathcal{G} = (N, E)$ where $N = \N_n$ is the set of nodes and
$E \subseteq [N]^2$ the set of edges linking the nodes.
Here, $[N]^2$ denotes the set of all two-element subsets of $N$.
Node $j$ is a neighbour of node $i$, if there is an edge $\set{i, j} \in E$.
The set of all $n_j$ neighbours of node $j$ is $N_j = \{i \mid i \in N, \set{i, j} \in E \}$ with $n_j = |N_j|$.


\section{Model of interconnected \acsp{mg}}
\label{sec:interConnectedMg}
In this section, we introduce a control-oriented discrete-time model of a network of interconnected \acp{mg}.
The model is motivated by \citet{HanBraRai17,Han2021} and considers $n\in\N$ \acp{mg} that are linked by an electrical grid (see \cref{fig:interconnectedMgDiag}).
Each $\ac{mg}_j$, $j \in \N_n$, is connected to a bus in the grid via a \ac{pcc} which allows for power exchange with others.
Despite the connection, each \ac{mg} is assumed to be always capable of running in island mode without power exchange via the~\ac{pcc}.

\begin{figure}
	\centering
	%

\newcommand{\lnwidth}{0.5pt}
\newcommand{\lngrid}{0.100pt}

\newcommand{\wInPlot}[2]{\ensuremath{w_{\mathrm{#1},#2}}}
\newcommand{\wDPlot}[1]{\wInPlot{d}{#1}}
\newcommand{\wRPlot}[1]{\wInPlot{r}{#1}}

\newcommand{\xSPlot}[1]{\ensuremath{x_{#1}}}
\newcommand{\pInPlot}[2]{\ensuremath{p_{\mathrm{#1},#2}}}
\newcommand{\pSPlot}[1]{\pInPlot{s}{#1}}
\newcommand{\pTPlot}[1]{\pInPlot{t}{#1}}
\newcommand{\pRPlot}[1]{\pInPlot{r}{#1}}
\newcommand{\pGPlot}[1]{\pInPlot{g}{#1}}

\newcommand{\scaleSymbols}{0.03}
\input{./figures/NewNodeCommands.tex}

\tikzstyle{grid} = [>=latex, line width=\lnwidth, color=black, font=\footnotesize]

\tikzset{almostMidArrow/.style={
		grid,
		draw,%
		decoration={%
			markings,%
			mark=at position 0.65 with \arrow{stealth},%
		},%
		postaction=decorate}}%

\tikzset{midArrow/.style={
		grid,
		draw,%
		decoration={%
			markings,%
			mark=at position 0.5 with \arrow{stealth},%
		},%
		postaction=decorate}}%
\tikzset{endArrow/.style={
		grid,
		draw,%
		decoration={%
			markings,%
			mark=at position 0.85 with \arrow{stealth},%
		},%
		postaction=decorate}}%
\tikzset{invS/.style={rectangle, draw=black, inner sep=0.1pt, line width=\lnwidth}}
\tikzset{plotNode/.style={rectangle, fill = none, draw = none, inner sep=0.1pt, line width=\lnwidth}}

\begin{tikzpicture}[xscale = 0.5, yscale = 0.55]

\draw[midArrow] (2.7*-1.5, 3-1.45) to (2.7*-1.5, 2.75-1.45) to node[grid, above] (yItoII) {$\phantom{as} p_{12}$} (2.7*1.75, 2.75-1.45) node[grid, left] {} to (2.7*1.75, 3-1.45); 
\draw[endArrow]  (2.7*-1.75, 3-1.45) to (2.7*-1.75, 2.25-1.45) to node[grid, right, yshift = 2] (yItoIII) {} (2.7*2, -2.25+1.45) node[grid, xshift = -0.5cm, yshift = 0.3cm]{$p_{13}$} to (2.7*2, -3+1.45); 
\draw[midArrow] (2.7*-2, 3-1.45) to node[grid, left] (yItoIV) {$p_{14}$} (2.7*-2, -3+1.45); 
\draw[midArrow] (2.7*1.5, -3+1.45) to (2.7*1.5, -2.75+1.45) to node[grid, below] (yIIItoIV) {$p_{34}$} (2.7*-1.5, -2.75+1.45) node[grid, right] {} to (2.7*-1.5, -3+1.45); 

\draw[grid, -] (2.7*-1.5 + 0.2, 3-1.45) to (2.7*-2.0 - 0.2, 3-1.45) node[grid, yshift=-3]{\tiny 1}; 
\draw[grid, -] (2.7*1.7 - 0.2, 3-1.45) to (2.7*1.8 + 0.2, 3-1.45) node[grid, yshift=-3]{\tiny 2}; 
\draw[grid, -] (2.7*1.5 - 0.2, -3+1.45) to (2.7*2 + 0.2, -3+1.45) node[grid, yshift=3]{\tiny 3}; 
\draw[grid, -] (2.7*-1.5+0.2, -3+1.45) to (2.7*-2.0 - 0.2, -3+1.45) node[grid, yshift=3]{\tiny 4}; 

\path[almostMidArrow] (2.7*-1.75, 3-1.45) -- node[left, yshift = 0.5mm]{$p_{1,p}$} (2.7*-1.75, 3.6-1.25)
	coordinate[xshift = 0] (p1c)
	coordinate[xshift = 30] (p1t)
	coordinate[xshift = -30] (p1b);
\path[almostMidArrow] (2.7*1.75, 3-1.45) -- node[left, yshift = 0.5mm]{$p_{2,p}$} (2.7*1.75, 3.6-1.25)
	coordinate (p2c)
	coordinate[xshift = 30] (p2t)
	coordinate[xshift = -30] (p2b);
\path[almostMidArrow] (2.7*1.75, -3+1.45) -- node[left, yshift = -0.5mm]{$p_{3,p}$} (2.7*1.75, -3.6+1.25)
	coordinate[xshift = 0] (p3c)
	coordinate[xshift = 30] (p3t)
	coordinate[xshift = -30] (p3b);
\path[almostMidArrow] (2.7*-1.75, -3+1.45) -- node[left, yshift = -0.5mm]{$p_{4,p}$} (2.7*-1.75, -3.6+1.25)
	coordinate[xshift = 0] (p4c)
	coordinate[xshift = 30] (p4t)
	coordinate[xshift = -30] (p4b);

\node[invS, above of = p1t, node distance = 7mm] (p1storage) {\GraphTristateS}
	node[grid, right of = p1storage, node distance = 0.65cm] {};
\node[invS, above of = p1c, node distance = 7mm] (p1res) {\GraphTristateR};
\node[invS, above of = p1b, node distance = 7mm] (p1thermal) {\GraphTristateT};

\draw[grid, -] (p1t)++(0.2, 0) -- (p1b) -- ++(-0.2, 0);
\draw[almostMidArrow] (p1storage) -- node[right]{$p_{1,s}$} (p1t);
\draw[almostMidArrow] (p1res) -- node[right]{$p_{1,r}$} (p1c);
\draw[almostMidArrow] (p1thermal) -- node[right]{$p_{1,c}$} (p1b);

\draw[grid, ->] (p1t) -- ++(0, -0.4) node[right] {$w_{1,l}$};

\draw [dashed]	($(p1storage.north) + (1.4, 0.1)$) --
				($(p1thermal.north) + (-0.7, 0.1)$) --
				($(p1b) + (-0.7, -0.7)$) node[below, xshift = 10, yshift = 2]{\tiny MG 1} --
				($(p1t) + (1.4, -0.7)$) --
				($(p1storage.north) + (1.4, 0.1)$);

\node[invS, above of = p2t, node distance = 7mm] (p2storage) {\GraphTristateS}
	node[grid, right of = p2storage, node distance = 0.55cm] {};
\node[invS, above of = p2c, node distance = 7mm] (p2res) {\GraphTristateR};
\node[invS, above of = p2b, node distance = 7mm] (p2thermal) {\GraphTristateT};

\draw[grid, -] (p2t)++(0.2,0) -- (p2b) -- ++(-0.2,0);
\draw[almostMidArrow] (p2storage) -- node[right]{$p_{2,s}$} (p2t);
\draw[almostMidArrow] (p2res) -- node[right]{$p_{2,r}$} (p2c);
\draw[almostMidArrow] (p2thermal) -- node[right]{$p_{2,c}$} (p2b);

\draw[grid, ->] (p2t) -- ++(0, -0.4) node[right] {$w_{2,l}$};

\draw [dashed]	($(p2storage.north) + (1.4, 0.1)$) --
				($(p2thermal.north) + (-0.7, 0.1)$) --
				($(p2b) + (-0.7, -0.7)$) --
				($(p2t) + (1.4, -0.7)$) node[below, xshift = -10, yshift = 2]{\tiny MG 2} --
				($(p2storage.north) + (1.4, 0.1)$);

\node[invS, below of = p3t, node distance = 7mm] (p3storage) {\GraphTristateS}
	node[grid, right of = p3storage, node distance = 0.55cm] {};
\node[invS, below of = p3c, node distance = 7mm] (p3res) {\GraphTristatePV};
\node[invS, below of = p3b, node distance = 7mm] (p3thermal) {\GraphTristateT};

\draw[grid, -] (p3t)++(0.2,0) -- (p3b) -- ++(-0.2,0);
\draw[almostMidArrow] (p3storage) -- node[right]{$p_{3,s}$} (p3t);
\draw[almostMidArrow] (p3res) -- node[right]{$p_{3,r}$} (p3c);
\draw[almostMidArrow] (p3thermal) -- node[right]{$p_{3,c}$} (p3b);

\draw[grid, ->] (p3t) -- ++(0, 0.4) node[right] {$w_{3,l}$};

\draw [dashed]	($(p3storage.south) + (1.4, -0.1)$) --
				($(p3thermal.south) + (-0.7, -0.1)$) --
				($(p3b) + (-0.7, 0.7)$) --
				($(p3t) + (1.4, 0.7)$) node[above, xshift = -10, yshift = -2]{\tiny MG 3} --
				($(p3storage.south) + (1.4, -0.1)$);

\node[invS, below of = p4t, node distance = 7mm] (p4storage) {\GraphTristateS}
	node[grid, right of = p4storage, node distance = 0.55cm] {};
\node[invS, below of = p4c, node distance = 7mm] (p4res) {\GraphTristatePV};
\node[invS, below of = p4b, node distance = 7mm] (p4thermal) {\GraphTristateT};

\draw[grid, -] (p4t)++(0.2,0) -- (p4b) -- ++(-0.2,0);
\draw[almostMidArrow] (p4storage) -- node[right]{$p_{4,s}$} (p4t);
\draw[almostMidArrow] (p4res) -- node[right]{$p_{4,r}$} (p4c);
\draw[almostMidArrow] (p4thermal) -- node[right]{$p_{4,c}$} (p4b);

\draw[grid, ->] (p4t) -- ++(0, 0.4) node[right] {$w_{4,l}$};

\draw [dashed]	($(p4storage.south) + (1.4, -0.1)$) --
				($(p4thermal.south) + (-0.7, -0.1)$) --
				($(p4b) + (-0.7, 0.7)$) node[above, xshift = 10, yshift = -2]{\tiny MG 4} --
				($(p4t) + (1.4, 0.7)$) --
				($(p4storage.south) + (1.4, -0.1)$);

\end{tikzpicture}%
%

	\caption{Topology of four interconnected \acp{mg} introduced by \citet{HanBraRai17}.}
	\label{fig:interconnectedMgDiag}
\end{figure}

\subsection{Model of a single microgrid}
\label{sec:interConnectedMg:singleMg}
The presented mathematical model of an \ac{mg} is composed of unit constraints, energy dynamics and constraints that account for a local power balance.
Note that the control scheme derived in this paper can be easily modified to work with alternative \ac{mg} configurations, such as, \acp{mg} with controllable loads or \acp{mg} in which renewable infeed cannot be curtailed.

Each $\ac{mg}_{j}$ is composed of $g_j\in\N$ \acp{dg} and $l_j\in\N$ loads.
A \ac{dg} can be a conventional, a storage or a renewable unit.
Unit $i \in \N_{g_j}$ is labelled $\ac{dg}_{j, i}$ such that
we can define the index set of conventional units as
${C_j = \{ i \mid i \in \N_{g_j}, \ac{dg}_{j,i} \text{ is a conventional unit} \}}$.
Similarly,
$S_j$ is the index set of storage units,
$R_j$ the index set of \ac{res} and
$L_j = \{g_j+1, \ldots, g_j + l_j\}\subset\N$ the index set of loads.
The number of conventional, storage and renewable units is $|C_j|$, $|S_j|$ and $|R_j|$ with $g_j = |C_j| + |S_j| + |R_j|$.

\subsubsection{Renewable energy sources}

\ac{res} like photovoltaic generators or wind turbines are assumed to be the predominant energy providers in the \acp{mg}.
The power from these units is uncertain as their availability depends on intermittent weather conditions.
At time instant $k\in\N^0$, the available power of unit ${i \in R_j}$ is denoted $w_{j, i}\ok$.
The power infeed $p_{j, i}\ok$ of this unit is limited by the available power $w_{j, i}\ok$, which, in turn, is bounded by the rated power $\overline{p}_{j, i}\in\R_{\geq 0}$, i.e.,
  \begin{equation} \label{eq:model:res}
    0 \leq p_{j, i}\ok \leq  w_{j, i}\ok \leq \overline{p}_{j, i},
		 \quad \forall i \in R_{j}.
  \end{equation}

\subsubsection{Storage units}\label{sec:model:storage}
\begin{subequations}\label{eq:model:storage}
We consider grid-forming storage units that allow an operation where all conventional units are temporarily disabled.
Stored energy and power of unit $i \in S_j$ are denoted $x_{j,i}\ok$ and $p_{j,i}\ok$, respectively.
The dynamics depend on whether a storage is charging or discharging.
Consider a variable $\delta_{j, i}\ok \in \{0, 1\}$, which is $1$ if unit $i$ is charging and $0$ if it is discharging.
With power limits $\underline{p}_{j,i}\in\R_{< 0}$ and $\overline{p}_{j,i}\in\R_{> 0}$ this relation can be implemented by
\begin{equation}\label{eq:model:storage:a}
  \underline{p}_{j,i} \delta_{j, i}\ok \leq p_{j, i}\ok \leq
	(1 - \delta_{j, i}\ok)\overline{p}_{j,i}, \quad \forall i \in S_j.
\end{equation}

With sampling time $T_{\text{s}} \in \R_{> 0}$ and storage efficiency $\eta_{j, i} \in (0, 1]$, the dynamics of storage unit $i$ read
\begin{multline}
	x_{j, i}(k + 1) = x_{j, i}\ok - \eta_{j, i}\delta_{j, i}\ok T_{\text{s}} p_{j, i}\ok \\
	  - \textstyle \frac{1}{\eta_{j, i}}(1 - \delta_{j, i}\ok) T_{\text{s}} p_{j, i}\ok, \quad \forall i \in S_j. \label{eq:model:storage:b}
\end{multline}
The energy is bounded by $\underline{x}_{j,i}\in\R_{\geq 0}$ and
$\overline{x}_{j,i}\in\R_{> 0}$, i.e.,
\begin{align}
  \underline{x}_{j,i} \leq x_{j, i}(k) \leq \overline{x}_{j,i}, \quad \forall i \in S_j. \label{eq:model:storage:c}
\end{align}
\end{subequations}

\begin{remark}
	Note that \eqref{eq:model:storage:b} comprises a nonlinear multiplication of decision variables $\delta_{j, i}\ok$ and $p_{j, i}\ok$.
	With the help of additional real-valued decision variables, \eqref{eq:model:storage:b} can be transformed into a set of affine constraints \citep{PRG2014}.
	Therefore, \eqref{eq:model:storage} can be used in \acp{miqcp}.
\end{remark}

\subsubsection{Conventional generators}
Typically, these units are used as a backup in times of low renewable infeed and little stored energy.
Let us represent the on/off condition of unit $i\in C_j$ by $\delta_{j, i} \in \{0, 1\}$ where
$\delta_{j, i} = 0$ means \enquote{off} and
$\delta_{j, i} = 1$ \enquote{on}.
This can be combined with the limits $\underline{p}_{j, i}\in\R_{> 0}$ and $\overline{p}_{j, i} \in\R_{> 0}$ into
\begin{equation}\label{eq:const:conv}
  \underline{p}_{j, i} \delta_{j, i}(k) \leq p_{j, i}(k) \leq \overline{p}_{j, i}\delta_{j, i}(k), \quad \forall i \in C_{j}.
\end{equation}

\subsubsection{Loads}
Uncontrollable electric demand is modelled as an uncertain input.
It is denoted by $w_{j, i}(k) \in \R_{\leq 0}$ for all $i \in L_j$.

\subsubsection{Point of common of coupling (PCC)}\acused{pcc}
The \ac{pcc} connects $\ac{mg}_j$ with the transmission network and thus allows for power exchange with other \acp{mg}.
Let us denote this power exchange by $p_{j, p}(k) \in \R$.
If $p_{j, p}(k) < 0$, then power is provided by $\ac{mg}_j$,
if ${p_{j, p}(k) > 0}$, then power is drawn by $\ac{mg}_j$.
This power is constrained by $\underline{p}_{j,p}\in\R_{< 0}$ and $\overline{p}_{j,p}\in\R_{> 0}$, i.e.,
\begin{subequations}\label{eq:power:pccAndBalance}
	\begin{equation}
		\label{eq:power:pcc}
		\underline{p}_{j, p} \leq p_{j,p}(k) \leq \overline{p}_{j, p}.
	\end{equation}
The local power balance of $\ac{mg}_j$ can be expressed by
	\begin{equation}
	  \label{eq:power:balance}
	  p_{j,p}\ok + \smallSum_{i = 1}^{g_j}p_{j, i}\ok + \smallSum_{i = g_j+1}^{g_j+l_j} w_{j, i}\ok = 0.
	\end{equation}
\end{subequations}

\subsection{Power exchange over the electric grid}
\label{sec:interconnectedMgs:powerExchange}

The transmission network is modelled as an undirected graph $\mathcal{G} = (N, E)$, where each node in $N$ represents an \ac{mg}, i.e., $|N| = n$.
Each element in $E$ represents a power line that connects two nodes, i.e., two \acp{mg}.
Let us denote the power exchange between two connected nodes oriented from $j\in N$ to $m\in N_j$ as $p_{jm}(k)$.

The power provided or consumed by $\ac{mg}_j$ via the \ac{pcc} is linked to the power exchange with its neighbours by
\begin{equation}\label{eq:node:power}
p_{j,p}\ok = \smallSum_{ m \in N_j} p_{jm}\ok, \quad \forall j \in N.
\end{equation}
We assume that the transmission network is composed of power lines with reactance to resistance ratio much greater than one.
Therefore, line losses are assumed to be negligible \citep{OuaDagDes+15, ParWieKyn+17, AnaMaeOca+18}.
For lossless lines, we then have that $p_{jm}(k) = -p_{mj}(k)$ and that a global power balance of the form $\sum_{j \in N } p_{j,p}(k)= 0$ holds.


\section{Conditional cooperation (\acs{cc}\acused{cc}) \acs{mpc} for interconnected \acsp{mg}} \label{sec:optimalControlProblem}

In \ac{mpc}, control actions are found by solving a finite horizon optimization problem \citep{rawlings2009model}.
Typically, the optimal control actions associated with the first prediction step are applied to the system at each execution.
At the next time step, initial conditions and forecasts are updated and the procedure is repeated.

\subsection{Control objectives}
\label{sec:controlObjectives}
The control objective of $\ac{mg}_j$ is given by
\begin{multline}\label{eq:sumOfUnitCosts}
  \ell_{j}(p_j, \delta_j) = \smallSum_{i \in C_j} \ell_{j,c}(p_{j, i}, \delta_{j, i}) + \smallSum_{i \in R_j} \ell_{j, r}(p_{j, i})\\
  + \smallSum_{i \in S_j} \ell_{j, s}(p_{j, i}) + \ell_{j, p}(p_{j, p}),
\end{multline}
with
$p_j = [\vec(p_{j, i})_{i \in  \mathbb{N}_{g_j}}\tran ~ p_{j, p}]\tran$,
$\delta_j = \vec(\delta_{j, i})_{i \in C_j \cup S_j}$.
The costs of the conventional, storage and renewable units are $\ell_{j, c}(\cdot, \cdot)\in\R_{\geq 0}$, $ \ell_{j, s}(\cdot)\in\R_{\geq 0}$ and $\ell_{j, r}(\cdot)\in\R_{\geq 0}$.
The cost associated with power exchange is ${\ell_{j, p}(\cdot)\in\R}$.

\begin{subequations}
The objective function of conventional unit $i\in C_j$ is based on running costs.
These can be approximated by
\begin{equation}\label{eq:cost:conventional}
  \ell_{j, c}(p_{j,i}, \delta_{j, i}) = a_{j,i}\delta_{j,i} +  a_{j,i}^{\prime}p_{j,i} + a_{j,i}^{\prime \prime} p_{j,i}^2,
\end{equation}
with weights $a_{j,i} \in \R_{> 0}$, $a_{j,i}^{\prime} \in \R_{> 0}$ and $a_{j,i}^{\prime \prime} \in \R_{> 0}$ \citep{ZGY2009}.
Note that the square term allows to model a maximum cost efficiency at power values below $\overline{p}_{j, i}$.

It is desirable to maximise infeed of \ac{res}.
Therefore, $\ell_{j,r}(\cdot)$ penalises a limitation of available infeed $w_{j,i}$.
With weight $a_{j,i} \in \R_{> 0}$, it can be expressed for all $i\in R_j$ by
\begin{equation}\label{eq:cost:res}
\ell_{j,r}(p_{j,i}) = a_{j,i} (p_{j,i} - w_{j, i})^2.
\end{equation}

Charging or discharging at high power can have a negative impact on the ageing of batteries~\citep{ShiLiChu17}.
This is modelled for storage unit $i\in S_j$ with weight $a_{j, i} \in \R_{> 0}$ by
\begin{equation}\label{eq:cost:storage}
  \ell_{j,s}(p_{j,i}) = a_{j, i}p_{j, i}^2.
\end{equation}

With selling/buying price $a_{j, p} \in \R_{ > 0}$ and trading cost weight $a_{j, p}^{\prime} \in \R_{ > 0}$, the power exchange cost reads
\begin{equation}\label{eq:cost:pcc}
  \ell_{j, p}(p_{j,p}) = a_{j, p} p_{j, p} + a_{j, p}^{\prime}|p_{j, p}|.
\end{equation}
\end{subequations}

\subsection{Central \acs{cc} \acs{mpc}} \label{sec:centralControl}
We assume that all \acp{mg} are capable of running in island mode.
Therefore, motivated by \citet{WanHua17}, a condition is included which ensures that each \acs{mg}'s connected operation cost does not exceed its islanded operation cost.
Before posing different \ac{mpc} problems, we need to introduce the sets of feasible power and energy values.

Let us collect the forecasts of available renewable power and load in
$\hat{w}_j = \vec(\hat{w}_{j, i})_{i \in R_j \cup L_j}$.
With this, the set of feasible power values of $\ac{mg}_j$ can be formulated as%
\begin{multline*}
  \mathcal{P}_j(\hat{w}_j, \delta_j) = \big\{ p_j \in \R^{g_j + 1} |
    \text{ eqs. \eqref{eq:model:res}, \eqref{eq:model:storage:a},
    \eqref{eq:const:conv}, \eqref{eq:power:pccAndBalance}} \\
    \text{hold and } w_j = \hat{w}_j
  \big\}.
\end{multline*}
Let us further define the vectors
$x_j\ok = \vec(x_{j,i}\ok)_{i \in S_j}$,
$p_{j, s}\ok = \vec(p_{j,i}\ok)_{i \in S_j}$,
$\delta_{j, s}\ok = \vec(\delta_{j,i}\ok)_{i \in S_j}$ and
$b_{j}(\delta_{j, s}\ok) = \vec(\eta_{j, i}\delta_{j, i}\ok + \frac{1}{\eta_{j, i}} (1 - \delta_{j, i}\ok))_{i \in S_j}$.
For $\ac{mg}_j$, constraint \eqref{eq:model:storage:c}, can be captured by the set
\begin{align*}
  \mathcal{X}_{j} = \{ x_{j} \in \R^{|S_j|} \mid
  \underline{x}_{j,i} \leq x_{j, i} \leq \overline{x}_{j, i}, ~ \forall i \in S_j \}.
\end{align*}

\subsubsection{Condition for cooperation (islanded operation)}

The optimal islanded cost for each \ac{mg} is deduced by solving an \ac{mpc} problem with zero \ac{pcc} power.
At $\ac{mg}_j$, we collect the decision variables over prediction horizon $H \in \N$ in
$\mathbf{p}_j = [p_j(k| k) ~ \cdots ~ p_j(k + H | k)]$, and
${\boldSwitch_j = [{\delta_j(k| k)} ~ \cdots ~ \delta_j(k + H | k)]}$.
Here, $k + h | k$ refers to a prediction for time $k + h$, $h \in \N_{[0, H]}$, performed at $k$.

Consider a measurement $x^0_{j}$ and a forecast $\hat{w}_j$
over prediction horizon $H$.
The \ac{mpc} problem associated with the islanded operation of $\ac{mg}_j$ reads as follows.

\begin{problem}[Islanded \ac{mpc}] \label{prob:islanded}
  \begin{subequations}
    \[
      \min_{\mathbf{p}_j, \boldSwitch_j} V_j(\mathbf{p}_j, \boldSwitch_j)
    \]
    \[
      \text{with} \quad V_j(\mathbf{p}_j, \boldSwitch_j) =
      \smallSum_{h = 0}^{H} \gamma^{h}\ell_{j}(p_j(k + h | k), \delta_j(k + h | k))
    \]
    subject to
      \begin{align}
        x_{j}(k | k) &= x^0_{j},\label{eq:problem:initial:state}\\
        x_{j}(k + h + 1 | k ) &= x_{j}(k + h| k ) - \nonumber\\
        & \hspace{-8mm}\diag(b_{j}(\delta_{j, s}\okhk)) T_{\text{s}} p_{j, s}(k + h| k), \label{eq:problem:dynamics}\\
        x_{j}(k + h + 1 | k ) &\in \mathcal{X}_{j}, \label{eq:problem:battery}\\
        p_j(k + h | k )  &\in  \mathcal{P}_j(\hat{w}_j(k + h | k), \delta_j(k + h | k)), \label{eq:problem:power:unit}\\
        p_{j, p}(k + h | k) &= 0, \label{eq:problem:pcc:power}
      \end{align}
  \end{subequations}
  for all $h \in \N_{[0, H]}$ where $\gamma \in (0, 1]$ is a discount factor which is used to emphasize decisions in the near future.
\end{problem}

\begin{assumption}\label{rem:islanded:feasibility}
  We assume that \cref{prob:islanded} is always feasible.
  This is the case, if the load can be fully served by the conventional units and $x^0_{j} \in \mathcal{X}_{j}$.
\end{assumption}

It is desired that no \ac{mg} has a disadvantage from grid-connected over islanded operation.
Using the optimal islanded operation cost $V_j^{I}$, this can be encoded in
\[
  \mathcal{C}_j(\boldSwitch_j, V_j^{I}) = \\
    \{ \mathbf{p}_j \in \R^{(g_j+1) \times (H+1)} \mid V_j(\mathbf{p}_j, \boldSwitch_j) \leq V_j^{I} \}.
\]

\subsubsection{Conditional cooperation \acs{mpc}}

The \ac{mpc} problem for a network of interconnected \acp{mg} decides the power of each \ac{dg} in each \ac{mg} along with the power exchange between the \acp{mg}.
With $\mathbf{P} = \vec(\mathbf{p}_j)_{j\in N}$ and $\boldSwitch = \vec(\boldSwitch_j)_{j\in N}$, it reads as follows.

\begin{problem}[Central \ac{cc} \ac{mpc}]
  \label{prob:central}
    \[
      \min_{\mathbf{P}, \boldsymbol{\delta}}
      V(\mathbf{P}, \boldsymbol{\delta})
    \]
    \begin{equation}\label{eq:prob:central:cost}
      \text{with} \quad V(\mathbf{P}, \boldsymbol{\delta}) = \smallSum_{j \in N} V_j(\mathbf{p}_j, \boldsymbol{\delta}_j)
    \end{equation}
  \begin{subequations}\label{eq:prob:central:constraints}
    subject to \eqref{eq:problem:initial:state}--\eqref{eq:problem:power:unit} and
      \begin{align}
      0 &= \smallSum_{j \in N} p_{j, p}(k + h | k), \label{prob:network:const}\\
      \mathbf{p}_j &\in \mathcal{C}_j(\boldSwitch_j, V_j^{I}), \label{prob:constraint:cost}
      \end{align}
      for all $h \in \N_{[0, H]}$, $j \in N$.
  \end{subequations}
\end{problem}

\Cref{prob:constraint:cost} ensures that the cost of no \ac{mg} in interconnected operation exceeds the islanded cost.
\Cref{prob:central} can be cast as a \ac{miqcp}.
Its computational complexity increases heavily with the number of \acp{mg}.
One way to solve \cref{prob:central} in a distributed manner is by relaxing the binary variables \citep{HanBraRai17}.
This, however, does not always yield a feasible solution to the original problem.
As an alternative, we propose an algorithm that ensures feasibility and can be solved in a distributed manner.


\section{Feasible decomposition for solving \acs{cc} \acs{mpc}}
\label{sec:outerApproximation}

The main idea of the proposed feasible-decomposition-based (\acs{fd}-based\acused{fd}) algorithm is to iteratively solve a sequence of feasible intermediate problems such that the cost is non-increasing.
As stated in \cref{alg:feasible:decomposition}, first a subproblem with fixed binary variables is solved to update the \ac{pcc} power.
Then, a subproblem with fixed \ac{pcc} power is solved to update the binary variables.
This scheme is repeated until a termination criterion is met.

\subsection{Subproblem with fixed binary variables}
At iteration $q$, the fixed binaries of $\ac{mg}_j$ are denoted $\boldSwitch_j^{q}$.
With $\boldSwitch^{q} = \vec(\boldSwitch_j^{q})_{j\in N}$, we modify \cref{prob:central} as follows.

\begin{algorithm}[tb]
	\caption{Feasible decomposition for \acs{cc} \acs{mpc}}
	\label{alg:feasible:decomposition}
	{\bf Initialisation}: Find $\mathbf{P}^{1}$ and $\boldsymbol{\delta}^{1}$ by solving \cref{prob:islanded}.

	\textbf{for} $q = 1, \ldots, q^{\max}$
	\begin{enumerate}
		\item Solve \cref{prob:relaxed:central} for $\boldsymbol{\delta} = \boldsymbol{\delta}^{q}$ to obtain the optimal values $\widetilde{\mathbf{P}}^{q} = \vec(\widetilde{\mathbf{p}}_j^{q})_{j\in N}$ which include
		$\tilde{p}_{j, p}^{q}\okhk$ for all $\ac{mg}_j$.
		\item Check termination criterion: \newline \textbf{if} $V(\mathbf{P}^{q}, \boldSwitch^{q}) = V(\widetilde{\mathbf{P}}^{q}, \boldSwitch^{q})$ holds, \textbf{then} go to \textbf{end}.
		\item Solve \cref{prob:update:islanded} for all $\ac{mg}_j$ with fixed \ac{pcc} power $p_{j, p}\okhk = \tilde{p}_{j, p}^{q}\okhk$ to obtain the optimal values $\mathbf{P}^{q + 1}, \boldsymbol{\delta}^{q+1}$.\label{alg:feasible:decomposition:3}
	\end{enumerate}
	\textbf{end:} Output $\mathbf{P}^q$, $\boldsymbol{\delta}^q$.
\end{algorithm}

\begin{problem}[Modified \ac{cc} \ac{mpc}]
	\label{prob:relaxed:central}
		\begin{align*}
			\min_{\mathbf{P}}
			 \smallSum_{j \in N} V_j(\mathbf{p}_j, \boldsymbol{\delta}_j)
		\end{align*}
		subject to \eqref{eq:problem:initial:state}--\eqref{eq:problem:power:unit},~\eqref{eq:prob:central:constraints}
		for all $h \in \N_{[0, H]}$, $j \in N$ with
		 \begin{equation}\label{eq:prob:relaxed:central:binary}
			\boldsymbol{\delta} = \boldsymbol{\delta}^{q}.
		 \end{equation}
\end{problem}

Because of \eqref{eq:prob:relaxed:central:binary}, constraint~\eqref{prob:constraint:cost} becomes convex and \cref{prob:relaxed:central}
can be cast as a convex second order cone problem which can be solved efficiently using interior-point methods.
Note that at iteration $q = 1$, a feasible choice of $\boldsymbol{\delta}^{1}$ is obtained from \cref{prob:islanded}.
We will now use the optimal solution to \cref{prob:relaxed:central}, $\widetilde{\mathbf{P}}^{q}$, to formulate the subproblem for step \ref{alg:feasible:decomposition:3} in \cref{alg:feasible:decomposition}.

\subsection{Subproblem with fixed \acs{pcc} power}
\label{sec:outerApproximation:fixedPccPower}

In \cref{prob:central}, \eqref{prob:network:const} accounts for the coupling between the \acp{mg} using only the \ac{pcc} power of the different \acp{mg}.
Fixing the \ac{pcc} power at iteration $q$ to the optimal result from \cref{prob:relaxed:central}, which is part of $\widetilde{\mathbf{P}}^{q}$, enables a decomposition into $|N|$ subproblems.
In \cref{alg:feasible:decomposition}, we use the solutions to these subproblems, which can be stated as follows, to update the binary variables $\boldsymbol{\delta}^{q+1}$.

\begin{problem}[Binary update \ac{cc} \ac{mpc}] \label{prob:update:islanded}
    \[
      \min_{\mathbf{p}_j, \boldsymbol{\delta}_j} V_j(\mathbf{p}_j, \boldsymbol{\delta}_j)
    \]
	subject to \eqref{eq:problem:initial:state}--\eqref{eq:problem:power:unit} and
	\begin{align}
		p_{j, p}(k + h | k) = \tilde{p}_{j, p}^{q}(k + h| k),
		\label{eq:problem:pcc:update}
	\end{align}
	for all $h \in \N_{[0, H]}$.
\end{problem}

\begin{remark}\label{rem:complexity}
	In \cref{prob:update:islanded}, the \ac{pcc} power is fixed via \eqref{eq:problem:pcc:update}.
	Therefore, it can be solved for each \ac{mg} separately, which is beneficial in two ways:
	\begin{enumerate*}[(i)]
		\item It enables a decentralized solution on different computing nodes.
		\item To solve \cref{prob:update:islanded} for all \acp{mg}, $\sum_{j = 1}^{n} 2^{(H+1)(|C_j| + |S_j|)}$ possible combinations of binary variables need to be considered.
		For \cref{prob:central}, this number is $2^{\sum_{j =1}^{n}(H+1)(|C_j| + |S_j|)}$, which is typically much higher.
		For the example in \cref{sec:caseStudy}, this would be $2 \cdot 10^{31} $ combinations for \cref{prob:central} compared to $3 \cdot 10^8$ for \cref{prob:update:islanded}.
	\end{enumerate*}
\end{remark}

\subsection[Properties of FD-based algorithm]{Properties of FD-based algorithm}
\label{sec:feasibleDecomposition:properties}

\begin{proposition}[Monotonicity]\label{prop:nonIncreasing}
	Given
	$\mathbf{P}^{q}$ and $\boldsymbol{\delta}^{q}$,
	as well as the subsequent
	$\widetilde{\mathbf{P}}^{q}$ obtained by solving \cref{prob:relaxed:central} for $\boldsymbol{\delta}^{q}$.
	Also given $\mathbf{P}^{q + 1}$ and $\boldsymbol{\delta}^{q + 1}$, obtained by solving \cref{prob:update:islanded} using $\widetilde{\mathbf{P}}^{q}$.
	Then, $V(\mathbf{P}^{q + 1}, \boldsymbol{\delta}^{q + 1}) \leq V(\mathbf{P}^{q}, \boldsymbol{\delta}^{q})$.
\end{proposition}

\begin{proof}
	Because of the common constraints of \Cref{prob:relaxed:central,prob:update:islanded,prob:islanded} and the structure of \cref{alg:feasible:decomposition},
	$\mathbf{P}^{q}$ is always a feasible solution to \cref{prob:relaxed:central}.
	Therefore, $V(\widetilde{\mathbf{P}}^{q}, \boldSwitch^{q}) \leq V(\mathbf{P}^{q}, \boldSwitch^{q})$.
	Moreover, $(\widetilde{\mathbf{p}}_{j}^{q}, \boldSwitch_{j}^{q})$ is a feasible solution to \cref{prob:update:islanded}.
	Therefore, $V_j(\mathbf{p}_{j}^{q+1}, \boldSwitch_{j}^{q + 1}) \leq V_j(\widetilde{\mathbf{p}}_{j}^{q}, \boldSwitch_{j}^{q})$ and
	hence $V(\mathbf{P}^{q+1}, \boldSwitch^{q + 1}) \leq V(\mathbf{P}^{q}, \boldSwitch^{q})$.
	\hfill\ensuremath{\square}
\end{proof}

\begin{corollary}[Cost at termination]\label{prop:terminationCriterion}
    If \cref{prob:update:islanded} exhibits a unique optimal solution, then the cost at termination $V(\widetilde{\mathbf{P}}^{q}, \boldSwitch^{q})$ represents the best (not necessarily optimal) cost one can obtain with \cref{alg:feasible:decomposition}.
\end{corollary}

\begin{proof}
	For fixed $\boldSwitch^{q}$, the cost $V(\cdot, \boldSwitch^{q})$ subject to \eqref{eq:power:balance} is strongly convex.
	The solutions $\mathbf{P}^{q}$ and $\widetilde{\mathbf{P}}^{q}$ always satisfy \eqref{eq:power:balance}.
	Therefore, $V(\mathbf{P}^{q}, \boldSwitch^{q}) = V(\widetilde{\mathbf{P}}^{q}, \boldSwitch^{q}) \iff \mathbf{P}^{q} = \widetilde{\mathbf{P}}^{q}$.
	Thus, for $\widetilde{\mathbf{P}}^{q} = \mathbf{P}^{q}$ \cref{prob:update:islanded} yields the same $\boldSwitch^{q+1} = \boldSwitch^{q}$.
	Because of the monotonic decrease (see \cref{prop:nonIncreasing}),
	the result of \cref{alg:feasible:decomposition} does not change in subsequent iterations and can therefore be terminated.
	\hfill\ensuremath{\square}
\end{proof}

\begin{remark}\label{rem:multipleSolutions}
	The solution to \cref{prob:update:islanded} might not always be unique.
	If this is the case, then only one solution is used in \cref{alg:feasible:decomposition} and all others are discarded.
	Note that this does not affect \cref{prop:nonIncreasing}.
	Further note that the cost at termination in such a case is the best that one can obtain for the selected solution to \cref{prob:update:islanded}.
\end{remark}


\section{Distributed solution for \acs{fd}-based algorithm}
\label{sec:distributedOptimisation}

In \cref{alg:feasible:decomposition}, the binary variables are updated in a decentralised manner via \cref{prob:update:islanded}.
To obtain a fully distributed implementation of the overall algorithm, we also need to solve \cref{prob:relaxed:central} in a distributed manner.
This is desirable since changes in the structure or the objective of each \ac{mg} can be handled by adapting local optimization problems and do not need to be shared with others.
In this section, we provide an example of such a distributed scheme which is based the work of \citet{ChaDenZav15} and only requires communication between neighbouring \acp{mg}.
Note that the presented approach could be easily replaced by others, e.g., the one presented by \citet{BRAUN201810}.

As stated in \cref{sec:interconnectedMgs:powerExchange}, \eqref{prob:network:const} can be equivalently expressed by a combination of
$p_{j,p}\ok = \smallSum_{ m \in N_j} p_{jm}\ok$ for all $j\in N$ and
$p_{jm}\ok = -p_{mj}\ok$ for all ${j\in N}$, ${m\in N_j}$.
The latter expression can be alternatively posed as ${0 = p_{j, t}(k) + \hat{p}_{j, t}(k)}$ with
$p_{j, t}(k) = \vec(p_{jm}(k))_{m \in N_j}$ and
${\hat{p}_{j, t}(k) = \vec(p_{mj}(k))_{m \in N_j}}$.
Hence, with matrices
${\mathbf{p}_{j, t} =  [p_{j, t}(k | k) ~ \cdots ~ p_{j, t}(k+H | k)]}$
as well as
$\hat{\mathbf{p}}_{j,t} =  [\hat{p}_{j, t}(k | k) ~ \cdots ~ \hat{p}_{j, t}(k+H | k)]$
and
${\mathbf{p}_{t} = [\mathbf{p}_{1, t}\tran ~ \cdots ~ \mathbf{p}_{n, t}\tran]}$,
we can rewrite \cref{prob:relaxed:central} as follows.

\begin{problem}[Modified \ac{cc} \ac{mpc} with line power]
	\label{prob:relaxed:central:powerFlow}
	\begin{align*}
		\min_{\mathbf{P}, \mathbf{p}_{t}}
		 \smallSum_{j \in N} V_j(\mathbf{p}_j, \boldsymbol{\delta}_j)
	\end{align*}
	subject to \eqref{eq:problem:initial:state}--\eqref{eq:problem:power:unit},~\eqref{prob:constraint:cost}
	and
	\begin{subequations}
	\begin{align}
		p_{j,p}(k+h|k) &= \smallSum_{ m \in N_j} p_{jm}(k+h|k), \\
		0 &= \mathbf{p}_{j,t} + \hat{\mathbf{p}}_{j,t}, \label{eq:power:exchange}
	\end{align}
	\end{subequations}
	for all $h \in \N_{[0, H]}$, $j \in N$ with $\boldsymbol{\delta}_j = \boldsymbol{\delta}_j^{q}$.
\end{problem}

Let us introduce
$\mathbf{p}_{jm} = [p_{jm}(k | k) ~ \cdots ~ p_{jm}(k+H | k)]$ and
the Lagrangian variable $\lambda_{jm}(k)\in\R$ for all $\{j, m\} \in E$.
Let us further define
$\boldsymbol{\lambda}_j = [\lambda_j(k | k) ~ \cdots ~ \lambda_j(k + H| k)]$ with entries
$\lambda_j(k) = \vec(\lambda_{jm}(k))_{m \in N_j}$
for all $j \in N$.
With \eqref{eq:power:exchange} and fixed parameter $\rho > 0$, the local \ac{al} of \cref{prob:relaxed:central:powerFlow} can then be formulated as
\begin{multline}\label{eq:local:augmented:lagrangian}
	\hat{L}_{j, \rho}(\mathbf{p}_j, \mathbf{p}_{j, t}, \boldsymbol{\lambda}_j, \hat{\mathbf{p}}_{j, t}) =
	f_{j}(\mathbf{p}_j, \mathbf{p}_{j, t}) + \\
	 \langle \boldsymbol{\lambda}_{j},
	\mathbf{p}_{j, t} + \hat{\mathbf{p}}_{j, t} \rangle
	+ \nicefrac{\rho}{2} \|\mathbf{p}_{j, t} + \hat{\mathbf{p}}_{j, t}\|^2.
\end{multline}
This local \ac{al} can be used to find a distributed solution for \cref{prob:relaxed:central:powerFlow} along the lines of \citet{ChaDenZav15}.
At each iteration, the local \ac{al} \eqref{eq:local:augmented:lagrangian} is minimised for fixed $\boldsymbol{\lambda}_{j}$ and $\hat{\mathbf{p}}_{j, t}$ to update $\mathbf{p}_j$ and $\mathbf{p}_{j, t}$ at $\ac{mg}_j$.
Subsequently, $\boldsymbol{\lambda}_{j}$ and $\hat{\mathbf{p}}_{j, t}$ are updated based on local optimal solutions $\mathbf{p}_{m, t}$ from all neighbours $m \in \N_j$.
For termination, the primal and dual residuals can be evaluated based on the updates of $\boldsymbol{\lambda}_j$ and $\mathbf{p}_{j, t}$ \citep{ChaDenZav15}.

\begin{remark}\label{rem:localTerminationCriterion}
	The termination criterion in step (2) of \cref{alg:feasible:decomposition} is equivalent to \enquote{$V_j(\mathbf{p}^{q}_{j}, \boldSwitch^{q}_{j}) = V_{j}(\widetilde{\mathbf{p}}^{q}_{j}, \boldSwitch^{q}_{j})$ holds for all $\ac{mg}_j$} because of \eqref{eq:prob:central:cost} and the strongly convex nature of $V_j(\cdot, \boldSwitch_j^{q})$ subject to \eqref{eq:power:balance}.
	This enables a distributed implementation of the overall scheme.
	In such a scheme, converge information could be spread over the network using, e.g., a flooding algorithm.
\end{remark}


\newcommand{\mgref}[1]{\ac{mg}\,#1}

\section{Case study}
\label{sec:caseStudy}

\begin{table}[!b]
  \centering
  \caption{Parameters of $\ac{mg}_j$, $j \in \N_4$. Values motivated by \citet{HanBraRai17,JMK2012}.}
  \small
  \label{table:parameters}
  \begin{tabular}{@{~}ll@{~}}
    \midrule
    $\text{Parameter}$ & $\text{Value}$\\
    \cmidrule{1-2}
    $\vec(\underline{p}_{j,c})_{c\in C_j, j\in\N_4}$ & $[0.1 \quad 0.25 \quad 0.1 \quad 0.25]$\,$\unit{pu}$  \\[-0.12em]
    $\vec(\overline{p}_{j,c})_{c\in C_j, j\in\N_4}$ & $[0.8 \quad 1 \quad 0.8 \quad 1]$\,$\unit{pu}$ \\[-0.12em]
    $[\underline{p}_{j,p}\quad \overline{p}_{j,p}]$  & $[-1\quad 1]\,\unit{pu} \quad \forall s\in S_j$ \\[-0.12em]
    $[\underline{p}_{j,s}\quad \overline{p}_{j,s}]$  & $[-1\quad 1]\,\unit{pu} \quad \forall s\in S_j$ \\[-0.12em]
    $[\underline{p}_{j,r}\quad \overline{p}_{j,r}]$  & $[0\quad 2]\,\unit{pu} \quad \forall r\in R_j$ \\[-0.12em]
    $[\underline{x}_{j,s}\quad \overline{x}_{j,s}]$  & $[0\quad 6]\,\unit{pu\,h} \quad \forall s\in S_j$ \\[-0.12em]
    $\vec(\eta_{j, s})_{j\in\N_4}$ & $[0.95 \quad 0.9 \quad 0.95 \quad 0.9]$ \\
    \midrule
    $\vec(a_{j,c})_{c\in C_j, j\in\N_4}$ & $[1.21 \quad 1.22 \quad 1.23 \quad 1.23] \cdot 10^{-1} $ \\[-0.12em]
    $\vec(a_{j,c}^{\prime})_{c\in C_j, j\in\N_4}$ & $[1.53 \quad 1.54 \quad 1.54 \quad 1.55]\,\unitfrac{1}{pu}$ \\[-0.12em]
    $\vec(a_{j,c}^{\prime \prime})_{c\in C_j, j\in\N_4}$ & $[1.82 \quad 1.90 \quad 2.01 \quad 2.04] \cdot 10^{-2} \,\unitfrac{1}{pu^2} $ \\[-0.12em]
    $[a_{j,p} \quad  a_{j, p}^{\prime}]$ & $[0.35 \quad 0.1]\,\unitfrac{1}{pu}$\\[-0.12em]
    $a_{j,s}$ & $\unitfrac[0.1]{1}{pu^2} \quad \forall s\in S_j$ \\[-0.12em]
    $a_{j,r}$ & $\unitfrac[1]{1}{pu^2} \quad \forall r\in R_j$ \\
    \bottomrule
  \end{tabular}
\end{table}

In the case study, the topology in \cref{fig:interconnectedMgDiag} is used.
It is composed of four \acp{mg} that are connected by power lines.
Each \ac{mg} comprises a load, a renewable, a conventional and a storage unit.
Parameters and weights are summarised in \cref{table:parameters}.
Available renewable infeed was calculated from wind and irradiance measurements provided by \citet{ARM2009}.
For the load, a realistic pattern was generated based on real-world \ac{mg} data.
For the \ac{mpc}, a sampling rate of $T_{\text{s}} = \unit[30]{min}$,
a prediction horizon of \unit[6]{h}, i.e., $H = 12$, and
a persistence forecaster for the prediction of load and available renewable infeed were chosen motivated by \citet{HanBraRai17}.
The power over the transmission lines was calculated using the linear \ac{ac} power flow model from \citet{HanBraRai17}.
All simulations were carried out for $336$ time steps using MATLAB~R2018b
with YALMIP \citep{Lof2004} and Gurobi~9.0.2.

\subsection{Closed-loop simulations}

\begin{table}[t]
  \centering
  \caption{Closed-loop costs with different \ac{mpc} approaches.}
  \label{table:simulationResults}
  \small
  \begin{tabular}{@{~}l@{~}rrrrrr@{~}}
  \toprule
    &  & \multicolumn{1}{c}{(i)} & \multicolumn{1}{c}{(ii)} & \multicolumn{1}{c}{(iii)} & \multicolumn{1}{c}{(iv)} & \multicolumn{1}{c}{(v)}\\[-0.12em]
    &  & \multicolumn{1}{c}{Island.}  & \multicolumn{1}{c}{Centr.} & \multicolumn{1}{c}{Hie.\,dis.} & \multicolumn{1}{c}{Prob.~\ref{prob:central}} & \multicolumn{1}{c}{Alg.~\ref{alg:feasible:decomposition}} \\
  \midrule
  $\text{\ac{mg}}_1$  &  & 245.4 & 2.7    & 3.3   & 23.2  & 20.0    \\[-0.12em]
  $\text{\ac{mg}}_2$  &  & 230.2 & 121.4  & 116.7 & 149.4 & 139.5 \\[-0.12em]
  $\text{\ac{mg}}_3$  &  & 170.6 & 112.4  & 111.5 & 113.1 & 112.0 \\[-0.12em]
  $\text{\ac{mg}}_4$  &  & 227.0 & 119.0  & 135.6 & 126.9 & 144.6 \\
  \midrule
  Sum                 &  & 873.2 & 355.6  & 367.1 & 412.6 & 416.1 \\
  \bottomrule
  \end{tabular}
\end{table}

We performed closed-loop simulations with the following approaches:
\begin{enumerate*}[(i)]
  \item Islanded \ac{mpc} (\cref{prob:islanded}),
  \item central \ac{mpc} without self-interest \citep{HanBraRai17},
  \item hierarchical distributed \ac{mpc} without self-interest \citep{HanBraRai17},
  \item central \ac{cc} \ac{mpc} (\cref{prob:central}),
  and
  \item \ac{fd}-based \ac{cc} \ac{mpc} (\cref{alg:feasible:decomposition}).
\end{enumerate*}
The closed-loop cost determined via \eqref{eq:sumOfUnitCosts} is summarised in \cref{table:simulationResults}.
It can be observed, that the cost with the \ac{cc} \ac{mpc} (iv) is around \unit[16]{\%} higher than with (ii), which does not respect the \acp{mg}' self-interest.
Similarly, using \cref{alg:feasible:decomposition} is around \unit[13]{\%} more expensive than using (iii).
It appears that this increase is dominantly caused by \eqref{prob:constraint:cost} which ensures that no \ac{mg} has a disadvantage from power exchange.
Still, we observe a decrease of more than \unit[50]{\%} compared to islanded operation when using \cref{alg:feasible:decomposition}.
Moreover, it can be noted that using \cref{alg:feasible:decomposition} leads to slightly higher costs compared to solving \cref{prob:central}.
This, however, seems acceptable considering that \cref{alg:feasible:decomposition} is more scalable and can be implemented in a distributed manner.

\begin{figure}[b]
  \centering
  %

\pgfplotsset{linestyle sample/.style={%
   only marks,
     mark=*,
     every mark/.append style={
      {mark size=0.7pt},
      fill=vgLightBlue,
      draw=none,
      line width=0pt,
      opacity = 0.6}
   }
}

\newcommand{\numberbox}[1]{\parbox{1.65em}{\raggedleft (#1)}}
\begin{tikzpicture}[font=\scriptsize]

   \begin{axis}[
    myPlot,
    height = 3.35cm,
    width = 7cm,
    xlabel = {$V^I_1 - V_1(\mathbf{P}^\star, \boldsymbol{\delta}^\star)$},
    xmin = -5,
    xmax = 20,
    ymin = 0.6,
    ymax = 5,
    y axis line style={white},
    ytick style={white},
    yticklabel style={xshift = 2mm},
    ytick={1, 2, 3, 4},
    yticklabels={Centr. fr. [5] \numberbox{ii}, Hier. dis. [5] \numberbox{iii}, Prob.~2 \numberbox{iv}, Alg.~1 \numberbox{v}},
    colormap={my colormap}{
       color=(vgLightBlue)
       color=(vgRed)
       },
    ]

   \foreach \fileName [count=\ti from 0] in {%
      objFigure_central_cost_diff.csv,
      objFigure_hierarchical_cost_diff.csv,
      objFigure_two_stage_central_cost_diff.csv,
      objFigure_two_stage_FD_cost_diff.csv%
   }{
      \addplot [scatter, linestyle sample, point meta={ifthenelse(x<0, 1, 0)}, point meta min=0, point meta max=1,]
         table[x expr= \thisrow{mg_1}, y expr= (\thisrow{time}/336)*0.3 + 0.85 + \ti, col sep=comma]
            {tables/\fileName};
   }

   \coordinate (islanded) at (axis cs: 0, 5);
   \addplot[thick, color = black, densely dotted] coordinates {(0, 0) (0, 5)};

\end{axis}

\node [anchor = north east, color = vgRed, inner sep=5pt, yshift = 13pt] at (islanded) {\begin{tabular}{@{}r@{}} larger \\[-0.2em] cost \end{tabular}};
\node [anchor = north west, color = vgLightBlue, inner sep=5pt, yshift = 13pt] at (islanded) {\begin{tabular}{@{}l@{}}lower or equal cost \end{tabular}};

\end{tikzpicture}%
  \caption{Difference between optimal islanded cost and the optimal cost of different control approaches for $\ac{mg}_1$ for $336$ simulation steps.}
  \label{fig:costLowerZero}
\end{figure}

\begin{figure*}[t]
  \centering
  %

\colorlet{colorThermal}{vgDarkBlue}
\colorlet{colorStorage}{vgOrange}
\colorlet{colorRes}{vgGreen}
\colorlet{colorDemand}{vgLightBlue}
\colorlet{colorGrid}{vgRed}

\pgfplotsset{linestyleThermal stacked/.style = {%
  color=colorThermal,
  fill=colorThermal!20!white,
}}

\pgfplotsset{linestyleStoragePower stacked/.style = {%
  color=colorStorage,
  fill=colorStorage!20!white
}}

\pgfplotsset{linestyleGridPower stacked/.style = {%
  color=colorGrid,
  fill=colorGrid!20!white
}}

\pgfplotsset{linestyleRes stacked/.style = {%
  color=colorRes,
  fill=colorRes!20!white,
}}

\pgfplotsset{linestyleDemand stacked/.style = {%
  color=colorDemand,
  fill=colorDemand!20!white,
}}

\pgfplotsset{limit/.style = {const plot, color=black, dash pattern=on 1pt off 3pt on 3pt off 3pt, thin}}

\newcommand{\PlotLimit}[1]{%
  \addplot[limit, forget plot] plot coordinates{
    (1,#1)
    (7*48,#1)
  };
}

\pgfplotsset{resultsPlot/.style={%
		myPlot,
    height = 3.0cm,
		xtick = {0, 48, ..., 672},
		xticklabels = {0, 1, ..., 14},
		xmin = 0,
		xmax = 336,
		clip=false,
    width=7.7cm,
    legend columns=1,
    legend style={
      at={(1.03, 1.0)},
      anchor=north west,
      draw=none,
      fill=none,
      legend cell align=left,
      /tikz/every even column/.append style={column sep=0.3cm}
      },
  	title style={
  		at={(0.5, 1.15)},
  		anchor=north,
  	},
	}
}

\pgfplotsset{resultsPlotPower/.style={%
    resultsPlot,
    stack plots=y, stack negative=separate, area style, enlarge x limits=false,
    xlabel = {},
    ymin = -2,
    ymax = 2,
  },
}

\begin{tikzpicture}[font=\scriptsize]
\begin{axis}[%
  resultsPlotPower,
  xticklabels = {~},
  ylabel={Power in $\unit{pu}$},
  title = {(a) MG~1},
  ]

  \addplot [linestyleStoragePower stacked] table [x=time, y=storagePower, col sep=comma] {tables/objFigure_system_1.csv} \closedcycle;
  \addplot [linestyleGridPower stacked] table [x=time, y=pccPower, col sep=comma] {tables/objFigure_system_1.csv} \closedcycle;
  \addplot [linestyleRes stacked] table [x=time, y=renewablePower, col sep=comma] {tables/objFigure_system_1.csv} \closedcycle;
  \addplot [linestyleThermal stacked] table [x=time, y=thermalPower, col sep=comma] {tables/objFigure_system_1.csv} \closedcycle;
  \addplot [linestyleDemand stacked] table [x=time, y=loadPower, col sep=comma] {tables/objFigure_system_1.csv} \closedcycle;

\end{axis}

\begin{axis}[%
  shift={(7.8cm, 0cm)},
  resultsPlotPower,
  xticklabels = {~},
  title = {(b) MG~2},
  ]

  \addplot [linestyleStoragePower stacked] table [x=time, y=storagePower, col sep=comma] {tables/objFigure_system_2.csv} \closedcycle;
    \addlegendentry{Storage}
  \addplot [linestyleGridPower stacked] table [x=time, y=pccPower, col sep=comma] {tables/objFigure_system_2.csv} \closedcycle;
    \addlegendentry{Grid}
  \addplot [linestyleRes stacked] table [x=time, y=renewablePower, col sep=comma] {tables/objFigure_system_2.csv} \closedcycle;
    \addlegendentry{RES}
  \addplot [linestyleThermal stacked] table [x=time, y=thermalPower, col sep=comma] {tables/objFigure_system_2.csv} \closedcycle;
    \addlegendentry{Convent.}
  \addplot [linestyleDemand stacked] table [x=time, y=loadPower, col sep=comma] {tables/objFigure_system_2.csv} \closedcycle;
    \addlegendentry{Load}

\end{axis}

\begin{axis}[%
  shift={(7.8cm, -2.2cm)},
  resultsPlotPower,
  xticklabels = {~},
  title = {(d) MG~3},
  ]

  \addplot [linestyleStoragePower stacked] table [x=time, y=storagePower, col sep=comma] {tables/objFigure_system_3.csv} \closedcycle;
  \addplot [linestyleGridPower stacked] table [x=time, y=pccPower, col sep=comma] {tables/objFigure_system_3.csv} \closedcycle;
  \addplot [linestyleRes stacked] table [x=time, y=renewablePower, col sep=comma] {tables/objFigure_system_3.csv} \closedcycle;
  \addplot [linestyleThermal stacked] table [x=time, y=thermalPower, col sep=comma] {tables/objFigure_system_3.csv} \closedcycle;
  \addplot [linestyleDemand stacked] table [x=time, y=loadPower, col sep=comma] {tables/objFigure_system_3.csv} \closedcycle;

\end{axis}

\begin{axis}[%
  shift={(0cm, -2.2cm)},
  resultsPlotPower,
  xticklabels = {~},
  ylabel={Power in $\unit{pu}$},
  title = {(c) MG~4},
  ]

  \addplot [linestyleStoragePower stacked] table [x=time, y=storagePower, col sep=comma] {tables/objFigure_system_4.csv} \closedcycle;
  \addplot [linestyleGridPower stacked] table [x=time, y=pccPower, col sep=comma] {tables/objFigure_system_4.csv} \closedcycle;
  \addplot [linestyleRes stacked] table [x=time, y=renewablePower, col sep=comma] {tables/objFigure_system_4.csv} \closedcycle;
  \addplot [linestyleThermal stacked] table [x=time, y=thermalPower, col sep=comma] {tables/objFigure_system_4.csv} \closedcycle;
  \addplot [linestyleDemand stacked] table [x=time, y=loadPower, col sep=comma] {tables/objFigure_system_4.csv} \closedcycle;

\end{axis}

\begin{axis}[%
  shift={(0cm, -4.4cm)},
  resultsPlot,
  height = 3.0cm,
  ylabel = {Energy in $\unit{pu\,h}$},
  title = {(e) Stored energy},
  ymin = 0,
  ymax = 6,
  ytick = {0, 2, 4, 6},
  xlabel = {Time in $\unit{d}$},
  ]


  \addplot[color = vgOrange] table [x=time, y=storedEnergy, col sep=comma]{tables/objFigure_system_1.csv};
  \addplot[color = vgRed] table [x=time, y=storedEnergy, col sep=comma]{tables/objFigure_system_2.csv};
  \addplot[color = vgGreen] table [x=time, y=storedEnergy, col sep=comma]{tables/objFigure_system_3.csv};
  \addplot[color = vgDarkBlue] table [x=time, y=storedEnergy, col sep=comma]{tables/objFigure_system_4.csv};

\end{axis}

\begin{axis}[%
  shift={(7.8cm, -4.4cm)},
  resultsPlot,
  height = 3.0cm,
  ylabel = {Power in $\unit{pu}$},
  title = {(f) Grid power},
  ymin = -1,
  ymax = 1,
  ytick = {-1, 0, 1},
  xlabel = {Time in $\unit{d}$},
  ]

  \addplot[color = vgOrange] table [x=time, y=line12, col sep=comma]{tables/objFigure_network_line.csv};
  \addplot[color = vgRed] table [x=time, y=line13, col sep=comma]{tables/objFigure_network_line.csv};
  \addplot[color = vgGreen] table [x=time, y=line14, col sep=comma]{tables/objFigure_network_line.csv};
  \addplot[color = vgDarkBlue] table [x=time, y=line34, col sep=comma]{tables/objFigure_network_line.csv};

  \addlegendimage{line legend, color = vgOrange}
  \addlegendentry{$x_1$ \& $p_{12}$}
  \addlegendimage{line legend, color = vgRed}
  \addlegendentry{$x_2$ \& $p_{13}$}
  \addlegendimage{line legend, color = vgGreen}
  \addlegendentry{$x_3$ \& $p_{14}$}
  \addlegendimage{line legend, color = vgDarkBlue}
  \addlegendentry{$x_4$ \& $p_{34}$}
\end{axis}

\end{tikzpicture}%
  \caption{Closed-loop simulation results obtained with distributed \ac{cc} \ac{mpc} approach.}
  \label{fig:caseStudyResults}
\end{figure*}

In \cref{fig:costLowerZero}, the difference between optimal islanded costs and the costs obtained with approaches (ii)--(v) for $\ac{mg}_1$ is shown for all simulation steps.
It can be seen that approaches (ii) and (iii) sometimes lead to costs that are higher than in islanded operation.
In contrast, (iv) and (v) ensure that no \ac{mg} has a disadvantage from trading.

\Cref{fig:caseStudyResults} shows simulation results for \cref{alg:feasible:decomposition}.
$\ac{mg}_3$ and $\ac{mg}_4$ have photovoltaic based \ac{res} and therefore regularly receive power at night.
Overall, the available renewable power can be very well used.
Especially $\ac{mg}_1$ provides a lot of renewable energy (which would have been partly curtailed in island mode) to the other \acp{mg}.

\subsection{\acs{fd}-based algorithm}
\label{sec:caseStudy:fdBasedAlgorithm}

In this section, we assess convergence and open-loop costs of a fully distributed implementation of \cref{alg:feasible:decomposition} using the approach proposed by \citet{ChaDenZav15}.
In \cref{fig:histogramIteration}, it can be seen that the algorithm requires on average $2.8$ iterations to converge and rarely (in \unit[12.5]{\%} of the executions) requires more than $4$ iterations.
The solve time was less than \unit[90]{s} for all executions of \cref{alg:feasible:decomposition}, which represents a significant decrease compared to \cref{prob:central} which required \unit[316]{s} on average.

\begin{figure}[h]
  \centering
  %

\begin{tikzpicture}[font=\scriptsize]
  \begin{axis}[
    myPlot,
    height = 2.65 cm,
    width = 7.0cm,
    clip = false,
    xtick = {0, 1, 2, ..., 10},
    xmin = 0.5,
    xmax = 10.5,
    ymin = 0,
    ytick = {0, 40, 80},
    x label style={at={(1, -0.4)}, anchor=south west},
    ylabel= {\begin{tabular}{@{}c@{}}No. of \\[-0.2em] samples \end{tabular}},
    xlabel = {\begin{tabular}{@{}l@{}}No. of \\[-0.2em] iterations \end{tabular}},
    ybar,
    y axis line style={white},
    ytick style={white},
    y axis shift left = -4pt,
    ymajorgrids,
    major grid style=white,
    axis on top,
    nodes near coords,
    every node near coord/.style={color = vgLightBlue, fill=none},
    nodes near coords align={above},
    ]
    \addplot[fill=vgLightBlue, draw = none, bar width=2.2mm,] table[y = data1YData, col sep=comma] {tables/histValue_number_points.csv};
  \end{axis}
\end{tikzpicture}%
  \caption{Number of iterations required by \cref{alg:feasible:decomposition}.}
  \label{fig:histogramIteration}
\end{figure}

\begin{figure}
  \centering
  %

\begin{tikzpicture}[font=\scriptsize]
  \begin{semilogxaxis}[
    myPlot,
    height = 5.46cm,
    width = 7.3cm,
    xmin = 1e0,
    xmax = 2e3,
    xtick = {1e0, 1e1, 1e2, 1e3, 1e4},
    y axis line style={white},
    ytick style={white},
    yticklabel style={xshift = 2mm},
    ytick={1, 2, 3, 4, 5, 6, 7, 8, 9, 10, 11},
    yticklabels={Sum, $q = 10$, $q = 9$, $q = 8$, $q = 7$, $q = 6$, $q = 5$, $q = 4$, $q = 3$, $q = 2$, $q = 1$},
    ymax = 11.5,
    clip=false,
    xlabel = {\begin{tabular}{@{}l@{}}No. of \\[-0.2em] iterations \end{tabular}},
    x label style={at={(1, -0.15)}, anchor=south west},
    ]

    \addplot [linestyle boxplot] table[y = iterationYData, col sep=comma]
    {tables/iterationFigure_iteration.csv};

    \addplot [linestyle boxplot] table[y = 10thStageIterYData, col sep=comma]
    {tables/iterationFigure_iteration.csv};

    \addplot [linestyle boxplot] table[y = 9thStageIterYData, col sep=comma]
    {tables/iterationFigure_iteration.csv};

    \addplot [linestyle boxplot] table[y = 8thStageIterYData, col sep=comma]
    {tables/iterationFigure_iteration.csv};

    \addplot [linestyle boxplot] table[y = 7thStageIterYData, col sep=comma]
    {tables/iterationFigure_iteration.csv};

    \addplot [linestyle boxplot] table[y = 6thStageIterYData, col sep=comma]
    {tables/iterationFigure_iteration.csv};

    \addplot [linestyle boxplot] table[y = 5thStageIterYData, col sep=comma]
    {tables/iterationFigure_iteration.csv};

    \addplot [linestyle boxplot] table[y = 4thStageIterYData, col sep=comma]
    {tables/iterationFigure_iteration.csv};

    \addplot[linestyle boxplot] table [y = 3rdStageIterYData, col sep=comma]
     {tables/iterationFigure_iteration.csv};

    \addplot[linestyle boxplot] table [y = 2ndStageIterYData, col sep=comma]
    {tables/iterationFigure_iteration.csv};

    \addplot[linestyle boxplot] table [y = 1stStageIterYData, col sep=comma]
    {tables/iterationFigure_iteration.csv};
  \end{semilogxaxis}

\end{tikzpicture}%
  \caption{Boxplots of iterations required by the distributed \ac{al} algorithm at each outer iteration of \cref{alg:feasible:decomposition} and sum at each use of \cref{alg:feasible:decomposition}.
  Here, the large dots mark the median and the white areas the lower and upper quartiles.
  The lines are the whiskers and the small dots outliers.}
  \label{fig:iterationDistributedOA}
\end{figure}

The fully distributed implementation of the \ac{cc} \ac{mpc} includes nested iterations, i.e.,
outer iterations of \cref{alg:feasible:decomposition}
and inner iterations of a distributed \ac{al} algorithm used to solve \cref{prob:relaxed:central}.
This algorithm terminates if the primal and dual residuals are below $5\cdot10^{-3}$.
Here, $\mathbf{p}_{j, t}$ and $\boldsymbol{\lambda}_j$ from the previous outer iteration are used to warm-start the inner \ac{al} algorithm.
\Cref{fig:iterationDistributedOA} illustrates the number of inner iterations required at each outer iteration
as well as the sum over the entire execution of \cref{alg:feasible:decomposition}.
This sum is on average $140$ with the maximum being $1435$.
Given the relatively short solve times at each iteration, the overall solve time of the distributed \ac{cc} \ac{mpc} is still much lower than the one of \cref{prob:central} which underlines the applicability of our approach.

Similar convergence results were obtained for the IEEE 14 bus benchmark grid \citep{IEEE14Bus} with each bus having one \ac{mg} with a renewable, a conventional, a storage and a load connected, leading to a setup with $14$ \acp{mg}.
Here, \cref{alg:feasible:decomposition} required on average $3.9$ and at maximum $12$ outer iterations.
In sum, the inner \ac{al} algorithm required on average $418$ and at maximum $958$ iterations.


\section{Conclusions}
\label{sec:conclusions}

In this paper, we proposed a conditional cooperation \ac{mpc} strategy for the operation of interconnected \acp{mg}.
The strategy preserves the self-interest of each \ac{mg} by only trading power if this does not increase individual costs.
A feasible-decomposition-based algorithm was proposed to find a (not necessarily optimal) solution to the associated mixed-integer program.
The algorithm was posed in a distributed way by employing an \acl{al} approach.
A comprehensive case study illustrated that the presented scheme preserves the \acp{mg}' self-interests while enabling power exchange within the network.
Future work will focus on the inclusion of uncertainty, e.g., in a similar fashion as proposed by \citet{HofSamHanRaiHeiBos2019, HaSoRa+2019}.
Moreover, the performance of the algorithm as a function of the number of \acp{mg} and their individual setup shall be investigated.





\bibliographystyle{elsarticle-harv}
\bibliography{IEEEabrv,databasePaper}

\end{document}